\documentclass[final,1p,11pt]{elsarticle}

\usepackage{lineno}

\usepackage{%
epsfig,array,tabularx,epsfig,mathrsfs,
graphicx,rotating,ifthen,amsfonts,ragged2e,soul}
\PassOptionsToPackage{hyphens}{url}
\usepackage[hyphens]{url}
\usepackage{hyperref,listings,color,breakurl}

\hypersetup{
  colorlinks=true,
  linkcolor=blue,
  citecolor=blue,
  urlcolor=blue
}

\newcommand{\beq}{\begin{equation}}
\newcommand{\eeq}{\end{equation}}

\chardef\til=126

\definecolor{literal}{rgb}{0.58,0,0.82}
\newcommand{\literal}[1]{\texttt{\color{literal}#1}}

\begin{document}

\lstset{%
 backgroundcolor=\color{white},   
 basicstyle=\small\ttfamily,        
 breakatwhitespace=false,         
 breaklines=true,                 
 captionpos=b,                    
 commentstyle=\color[rgb]{0,0.6,0}, 
 escapeinside={\%*}{*)},          
 extendedchars=true,              
 keepspaces=true,                 
 frame=tb,
 keywordstyle=\color{blue},       
 rulecolor=\color{black},         
 showspaces=false,                
 showstringspaces=false,          
 showtabs=false,                  
 stepnumber=2,                    
 stringstyle=\color{literal}, 
 tabsize=2,                        
 title=\lstname,                   
 numberstyle=\footnotesize,
 basewidth={0.5em,0.5em},
 aboveskip=12pt, belowskip=0pt
}

\begin{frontmatter}

\title{
Cross-platform validation and analysis environment for particle physics
}

\author[add1]{S.V.~Chekanov\corref{cor1}}
\ead{chekanov@anl.gov}

\author[add2,add1]{I.~Pogrebnyak}
\ead{ivanp@msu.edu}

\author[add3]{D.~Wilbern}
\ead{dwilbern@nmu.edu}

\cortext[cor1]{Corresponding author}
\address[add1]{
HEP Division, Argonne National Laboratory,
9700 S. Cass Avenue,
Argonne, IL 60439, USA.
}
\address[add2]{
Department of Physics and Astronomy,
Michigan State University,
567 Wilson Road, East Lansing, MI 48824, USA.
}
\address[add3]{
Northern Michigan University,
1401 Presque Isle Avenue,
Marquette, MI 49855, USA.
}

\begin{abstract}
A multi-platform validation and analysis framework  for public Monte Carlo simulation
for high-energy particle collisions is discussed.
The front-end of this framework uses the Python programming language, while the back-end is written in Java,
which provides a multi-platform environment that can be run from a web browser and can easily be deployed at the grid sites.
The analysis package includes all major software tools used in high-energy physics,
such as Lorentz vectors, jet algorithms, histogram packages, graphic canvases, and tools
for providing data access.
This multi-platform software suite, designed to minimize OS-specific maintenance and deployment time, is used for online validation of
Monte Carlo event samples through a web interface.
\end{abstract}

\begin{keyword}
Monte Carlo, \sep format \sep IO \sep analysis software 
\PACS 07.05.Wr  \sep 07.05.Kf  \sep 07.05.-t 
\end{keyword}

\end{frontmatter}


\section{Introduction}

Development of data-analysis software is one of the most effort consuming components of nearly every project in high-energy physics.
Consequently, engineering of analysis packages which require minimum time for deployment,
maintenance, and porting to new platforms, is an important direction,
especially for long-term projects and projects without dedicated funding for computer support.
This is precisely the category that includes detector performance and physics studies for next-generation circular colliders.
Currently, these projects are community driven and often
do not have dedicated computer infrastructure needed for detailed exploration of physics using large samples of Monte Carlo (MC) events.
Therefore, a software environment  and deployment model that can help reduce
maintenance effort  presents an unambiguous advantage.
Conversely, the exploratory nature of future collider studies provides for
a good setting for testing of innovative computation models.


This paper discusses a validation and analysis framework that is built from the ground
to work with the HepSim data repository \cite{Chekanov:2014fga} 
that stores Monte Carlo predictions for HEP experiments. 
To ensure platform independence and support data processing using a web browsers,
the analysis environment runs on the Java platform, which also allows for the end-user analysis scripts to be written in Jython,
a Java implementation of Python.
This setup creates a familiar analysis environment, since Python is often used by the current LHC experiments,
and Java syntax is very similar to that of C++.
Therefore, such a framework has a potential to  target a larger audience within HEP community,
simplifying and facilitating exploration of physics for the next generation of collider experiments.

It should be pointed out that the same approach has been proposed for studies of $e^+e^-$ collisions
back in 2000, when the first version of FreeHep and AIDA (Abstract Interface for Data Analysis) was developed \cite{freehep}.
Unlike FreeHep  which is primary designed for linear $e^+e^-$ collider studies,
HepSim software is also used for circular colliders, including $pp$ colliders.
HepSim analysis framework is similar to {\tt PyROOT} and is fitting for Python programming, 
with classes named conveniently to reduce code verbosity.
As discussed in this paper, HepSim also includes jet algorithms which are popular for $pp$ collisions.

The choice of Java and Python for HepSim framework offers some key advantages, such as
\begin{enumerate}
\setlength{\itemsep}{1pt}
\setlength{\parskip}{0pt}
\setlength{\parsep}{0pt}
\item hardware, operating system, and library independence (except Java itself), making deployment trivial and effortless,
\item automatic memory management, eliminating frustratingly common sources of errors, and
\item simple, compressed, and lossless data storage format.
\end{enumerate}
These features eliminate all portability issues and give HepSim an edge in terms of programmer productivity, allowing researcher to concentrate more on physics and less on bookkeeping. Due to optimization provided by Java Virtual Machine employing just-in-time compilation technology, HepSim remains competitive in terms of data processing speed with natively compiled analysis software.

To access data from the HepSim Monte Carlo event catalog  \cite{Chekanov:2014fga},
one needs to download a small (25~MB) software package. The software requires Java version 7 or above.
The package includes Java classes for data access ({\tt browser\_promc.jar}), physics and graphical classes ({\tt hepsim.jar}), and a complete
Jython distribution ({\tt jython.jar}) with required modules ({\tt modules.jar}).
To minimize the package size for the online usage, only essential Python modules are used.
The software framework can be used on Windows computers, but here will will discuss Linux/Mac OS for simplicity.
Bash shell users can download and setup the framework by running the following commands:

\vspace{0.5cm}
\begin{lstlisting}[language=bash]
wget http://atlaswww.hep.anl.gov/hepsim/soft/hs-toolkit.tgz -O - | tar -xz
source hs-toolkit/setup.sh
\end{lstlisting}
The last command sets up the working environment. One can find HepSim commands by using the {\tt hs-help} command that prints this help message:

\begin{lstlisting}
HepSim toolkit version: 2
Commands:
=========
hs-find      - find a URL of a MC sample
hs-get       - download MC samples in multiple threads
hs-info      - validate the ProMC file or dump separate events
hs-meta      - read meta information
hs-ls        - list all files from a given data sample
hs-ide       - run an editor and process analysis *.py script
hs-run       - process analysis *.py script in a batch mode
hs-view      - view ProMC file from the HepSim database
hs-exec      - execute external commands
hs-extract   - extract N events from a file and write them to other file
hs-pileup    - create a new file mixing signal events with pile-up events
hs-distiller - convert ProMC file to zip64 format (distiller)
hs-help      - shows this help
\end{lstlisting}
As an example, one can run Python code using the command {\tt hs-ide test.py}, or in a batch mode using {\tt hs-run test.py}.

Alternatively, the framework can be used through a Java Web Start interface.
In this case, it uses a minimal version of the HepSim package with simplified Jython and
reduced sizes of Java jar files.

This paper discusses how to access Monte Carlo (MC) datasets stored in HepSim using the Python interface or Java,
manipulate data containers and perform a full scale analysis using local files, or streaming the files over the network.
Our discussion mainly focuses on $pp$ collision events, but the examples given
in this paper can also be used for other particle collision experiments.

\section{Data access}

The main Java package for data access is called {\tt hepsim.HepSim}.
It includes static Java classes for retrieving MC data lists from the HepSim catalog, and streaming
event records over the network (when reading data using URL).

Here is a simple Python script example to list HepSim MC files using the name of a dataset.

\begin{lstlisting}[language=Python]
from hepsim import HepSim
url   = HepSim.urlRedirector("tev100pp_ttbar_mg5")
flist = HepSim.getList(url)
print flist                         # print list of files for this dataset
\end{lstlisting}
Run this example as {\tt hs-run example.py}. Alternatively, use the editor {\tt hs-ide example.py}.
If you know the direct URL link to the MC dataset, use this example:
\begin{lstlisting}[language=Python]
from hepsim import HepSim
flist = HepSim.getList(url)         # url points to dataset
print flist
\end{lstlisting}

If MC datasets have been downloaded to the local file system as discussed in~\cite{Chekanov:2014fga}, one can build a list of files with:
\begin{lstlisting}[language=Python]
from jhplot.utils import FileList
flist = FileList.get("./data","promc")
\end{lstlisting}

Reading data typically means reading either of:
(1)~meta-data record, or the file header;
(2)~event record;
(3)~particle record.
This example shows how to read units stored in the header file.
\begin{lstlisting}[language=Python]
from proto import FileMC
f = FileMC("data.promc")            # input file from local directory or URL
print f.getDescription()            # get description
header = f.getHeader()              # to access information on this file
un=float(header.getMomentumUnit())  # conversion energy units
lunit=float(header.getLengthUnit()) # conversion length units
\end{lstlisting}

\section{Working with truth-level MC samples}

\subsection{The Lorentz vector class}

The framework includes a Lorentz vector class, {\tt LParticle}, similar to other implementations commonly used in HEP software.
The class {\tt LParticle} is an extended version of {\tt HepLorentzVector} developed by FreeHep~\cite{freehep}.
It provides methods to compute all necessary kinematic variables of a Lorentz particle, such as {\tt perp()} (transverse momentum),
{\tt rapidity()}, and {\tt pseudoRapidity()}.

Shown below is a typical example to construct Lorentz vectors from data stored in the  ProMC data format \cite{2013arXiv1306.6675C,2013arXiv1311.1229C} which uses
a 64-bit {\it varint} (varying-length integers) encoding that allows for very effective data storage and streaming of Monte Carlo event files over the network.
To access particles' 4-momenta, {\it varint} values need to be converted to floating point values, as shown in the example.

\begin{lstlisting}[language=Python]
from proto import FileMC from hephysics.particle import LParticle
f = FileMC("data.promc")            # input file from local directory or URL
un=float(header.getMomentumUnit())  # conversion energy units
lunit=float(header.getLengthUnit()) # conversion length units
for i in range(f.size()):           # run over all events in the file
  eve = f.read(i)                   # ProMCEvent record for event "i"
  pa = eve.getParticles()           # particle information
  for j in range(pa.getPxCount()):  # extract 4-momenta
    px,py,pz = pa.getPx(j)/un,pa.getPy(j)/un,pa.getPz(j)/un
    e = pa.getEnergy(j)/un          # energy
    m = pa.getMass(j)/un            # mass
    p = LParticle(px,py,pz,e,m)     # fill a Lorentz vector
\end{lstlisting}
The class {\tt LParticle} has all the required methods to work with a typical particle used in high-energy applications.
As an example, two particles can be added together to construct their invariant mass:
\begin{lstlisting}[language=Python]
gam1 = LParticle(px_1,py_1,pz_1,e_1)
gam2 = LParticle(px_2,py_2,pz_2,e_2)
gam1.add(gam2)                      # combine 2 photons, gam1 is now di-photon
mass=gam1.calcMass()                # calculate di-photon mass
print "pT=",gam1.perp(), " eta=",gam1.rapidity()
\end{lstlisting}

\subsection{Pileup re-weighting}

The HepSim analysis package includes a tool for mixing events from Monte Carlo events that contain
``signal'' events corresponding to the inelastic (minbias) processes.
Thus, the tool can be used to create realistic $pp$ collision datasets including ``pileup'' events.
Each signal event from the signal ProMC file is mixed with {\tt N} (supplied as an argument) randomly chosen events from the minbias file.
A minbias file with a large number of events should be used to minimize the reuse of the same events in the pileup-mixing.
If the {\tt N} argument is prefixed with the character {\tt p}, the number of minbias events that are mixed with signal events will be randomly sampled from a Poisson distribution.
Here is a usage example of this tool:

\begin{lstlisting}[language=bash]
hs-pileup p100 signal.promc minbias.promc output.promc
\end{lstlisting}

Here each signal event in \texttt{signal.promc} is mixed with a number of events from \texttt{minbias.promc}, according to a Poisson distribution, and the mixed events written to the \texttt{output.promc} file.
The signal and minbias sub-events can be distinguished by the value of the {\tt barcode} variable,
with {\tt barcode==0} corresponding to signal, and positive values enumerating minbias sub-events.

\subsection{Jet clustering algorithm}


A major goal of HepSim repository is to provide platform independent tools for analysis of Monte Carlo predictions for HEP experiments that can be run from a web browser on the client's side.
In order to meet this requirement, we provide an optimized Java implementation of generalized $k_\mathrm{t}$ jet clustering algorithms.

In recent years {\tt FastJet}~\cite{fastjet_man} has become a standard tool for jet clustering in HEP and heavy ion physics.
It is an extensively tested software package, providing $O(N^2)$ and $O(N\log N)$ complexity clustering strategies~\cite{fastjet_n3}.
{\tt FastJet} benchmarks~\cite{fastjet_unpub} show that clustering strategies with $O(N^2)$ complexity perform the fastest for events with numbers of particles characteristic for hadron and lepton collider experiments.
The more sophisticated, $O(N\log N)$ complexity, strategies provide advantage only for events with more than 10,000 particles, which arise only in experiments with heavy ion collisions.
This is because $O(N\log N)$ strategies, employing Voronoi diagrams~\cite{Aurenhammer}, require more computationally demanding bookkeeping, which, for low multiplicity events, takes longer to process than the time saved due to relative complexity reduction.
Consequently, in our Java jet clustering algorithm, \texttt{N2Jet}, we implement only $O(N^2)$ complexity strategies relevant for HEP Monte Carlo events for present and future collider experiments.
The adopted optimizations are described below.

\subsubsection{Generalized $k_\mathrm{t}$ algorithm}

The original $k_\mathrm{t}$ jet clustering algorithm~\cite{coxkt} is defined as follows.
\begin{enumerate}
  \item Find the value of a distance measure to the beam, $d_{i\mathrm{B}}$, for each (pseudo-)particle in the event,
  \begin{equation}
    d_{i\mathrm{B}} = k_\mathrm{t}^2.
  \end{equation}
  \item Find the value of a pairwise distance measure between particles, $d_{ij}$, for each (pseudo-)particle in the event,
  \begin{equation}
    d_{ij} = \min\left(k_{\mathrm{t}i}^2,k_{\mathrm{t}i}^2\right)R_{ij}^2, \quad
    R_{ij}^2 = \left(y_i-y_j\right)^2 + \left(\phi_i-\phi_j\right)^2.
  \end{equation}
  \item If the shortest distance is a $d_{i\mathrm{B}}$, identify the (pseudo-)particle as a jet and exclude it from further clusterization. If the shortest distance is a $d_{ij}$, then combine (pseudo-)particles' 4-momenta into a single pseudo-particle (pseudo-jet) and keep it in the event.
\end{enumerate}
Here, $i$ and $j$ are (pseudo-)particle indices, $k_\mathrm{t}$ is transverse momentum, and $y$ and $\phi$ are rapidity and azimuthal angle respectively.
The procedure is to be repeated until all the particles have been combined into jets.
The usual (re)combination scheme employed is 4-momentum addition, although other options may be desirable~\cite{fastjet_man}.
The former is called the $E$-scheme in {\tt FastJet} terminology.

The generalized $k_\mathrm{t}$ algorithm~\cite{genkt} is obtained by redefining the distance measures in the following way.
\begin{equation}
  d_{i\mathrm{B}} = k_\mathrm{t}^{2p}, \quad
  d_{ij} = \min\left(k_{\mathrm{t}i}^{2p},k_{\mathrm{t}i}^{2p}\right)\frac{R_{ij}^2}{R^2}, \quad
  R_{ij}^2 = \left(y_i-y_j\right)^2 + \left(\phi_i-\phi_j\right)^2,
\end{equation}
where $R$ is the jet radius parameter, and $p$ is the parameter governing the relative power of the energy versus geometrical ($R_{ij}$) scales.
The commonly used cases are:
\begin{itemize}
  \item $p= 1$ : $k_\mathrm{t}$ algorithm,
  \item $p= 0$ : Cambridge/Aachen algorithm,
  \item $p=-1$ : anti-$k_\mathrm{t}$ algorithm.
\end{itemize}

In our Java \texttt{N2Jet} algorithm, we implement the generalized $k_\mathrm{t}$ algorithm for ${p=\pm1,0}$, the 4-momentum addition recombination scheme, and $O(N^2)$ complexity optimization discussed below.

\subsubsection{Optimizations}
Geometric factorization is the most significant optimization,
reducing algorithm complexity from $O(N^3)$ to $O(N^2)$.
It is proved by the {\tt FastJet} Lemma~\cite{fastjet_n3}.
By the lemma, the shortest $d_{ij}$ in an event is found in the set of $d_{ij}$'s computed only with particles' geometric nearest neighbors.
The observation is intuitive, after realizing that the shortest distance $d_{ij}$ in the event must be formed with the shortest $R_{ij}$ for either $i$ or $j$, because otherwise either $k_{\mathrm{t}i}^2$ or $k_{\mathrm{t}j}^2$ can be used to form a shorter $d_{ik}$ or $d_{jk}$ with a closer particle $k$.
The computational optimization comes from storing only $d_{ij}$ and $j$ for every particle $i$ instead of storing all $d_{ij}$ values for every $(i,j)$ pair.
Thus, instead of storing (and reading) $N^2$ values (or $N(N-1)/2$, if repeated value are not stored) only $2N$ values are required.

Another optimization is obtained by subdividing the $y$--$\phi$ space into square tiles with half-width at least equal to the $R$ parameter. Then, if the nearest neighbor of particle $i$, found within the same tile, is closer than the adjacent tile, it must be the nearest over the whole space, and no more $R_{ij}$ values need to be computed. If that nearest neighbor is farther than any of the adjacent tiles, than only those tiles need to be checked for nearest neighbors. Searching further than the adjacent tiles is not needed, because $d_{i\mathrm{B}}$ is always shorter than $d_{ij}$ to those particles. To improve performance, tiling is only used with evens containing 50 particles or more, and is turned off when the number of particles decreases to 20.

The last optimization comes from storing all the (pseudo-)particle related values in a single class and using linked lists to provide means of iterating over them.
In comparison to use of arrays, this reduces random access insert/erase complexity from $O(N)$ to $O(1)$. In comparison to fixed size array implementation, link lists allow to reduce the number of loop iterations as the clusterization sequence progresses.

\subsubsection{Benchmark}
\begin{figure}[!htb]
  \centering
  \includegraphics[width=0.7\linewidth]{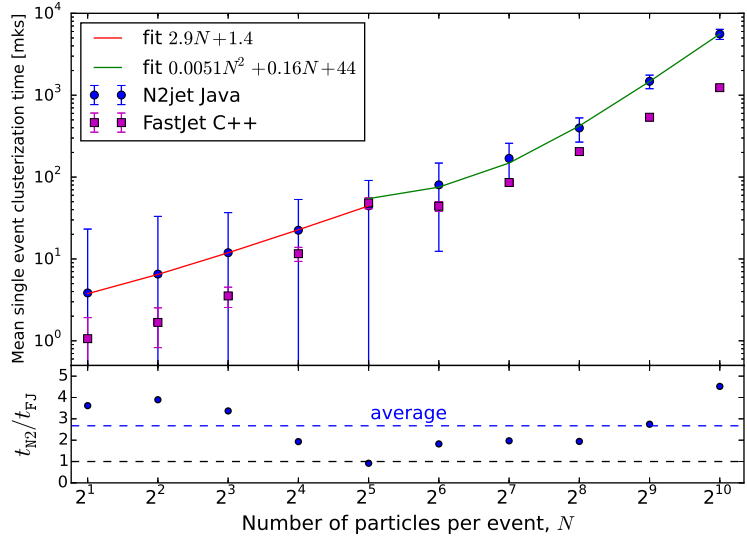}
  \caption{Benchmark comparing Java \texttt{N2Jet} and C++ \texttt{FastJet} performance
for AntiKt algorithm with $R=0.4$ using 100,000 events of each size.
The data points in the top panel show the mean computation time per single event. The error bars show standard deviation of the time per single event.
The bottom panel shows ratio between \texttt{N2Jet} and \texttt{FastJet} execution time, with the dashed line at 2.7 indicating the average.
The benchmark was conducted on an
Intel\textsuperscript{\textregistered} Core\texttrademark~i7-3632QM 2.2GHz processor.
}
  \label{fig:benchmark}
\end{figure}

A benchmark was conducted to compare Java \texttt{N2Jet} and \texttt{FastJet} performance. Events of different sizes with randomly generated particles were used. The particles' 4-momenta were generated with uniformly distributed $\phi$, $\eta$ distributed as the first half-period of cosine between -5 and 5, and $p_\mathrm{T}$ and mass distributed exponentially, with exponents of 0.007 and 0.05 respectively, and minimum $p_\mathrm{T}$ of 10 GeV. The benchmark results are are shown in Figure~\ref{fig:benchmark}.
On average, the Java implementation is 2.7 time slower than the C++ one.
There could be several reasons for this, such as a larger memory usage of high-level Java classes, or due to a concrete algorithmic implementation
that is different from the C++ code of {\tt FastJet}. The performance of \texttt{N2Jet} is satisfactory for the main goal -- 
validations of  typical simulated samples with several million $pp$ collision events using the Java Web start technology.
As can be seen for both \texttt{N2Jet} and \texttt{FastJet}, the algorithm's complexity is nearly linear for small events (memory allocation dominates), but for larger numbers of particles per event the complexity acquires characteristic quadratic dependence, as discussed previously.

\subsubsection{Application program interface}

To use the {\tt N2Jet} clustering algorithm one needs to call the respective class constructor, {\tt JetN2(R,alg,minPT)},
where \texttt{R} is the jet size, \texttt{alg} is the algorithm type, which can be
\literal{"antikt"}, \literal{"kt"}, or \literal{"ca"} (for the Cambridge/Aachen algorithm).
This change in the name from {\tt N2Jet} to {\tt JetN2(R,alg,minPT)} was needed to move the original stand-alone version
of this package to a version which is integrated with the rest of HepSim Java classes.
The string comparison is not case sensitive.
One can also pass an optional \texttt{minPT} argument, which specifies minimum $p_\mathrm{T}$ of constructed jets. Below is a typical example of jet construction using $pp$ collision events.
\begin{lstlisting}[language=Python]
from proto import FileMC from hephysics.hepsim import PromcUtil
from hephysics.jet import JetN2

clust = JetN2(0.5,"antikt",20)      # anti-kt jets with R=0.5 & min_pT=20
print clust.info()                  # print jet clusterization settings
f = FileMC("data.promc")            # input file from current directory / URL
header = f.getHeader()              # to access information on this file
un=float(header.getMomentumUnit())  # conversion energy units
lunit=float(header.getLengthUnit()) # conversion length units
for i in range(f.size()):           # run over all events in the file
    eve = f.read(i)                 # ProMCEvent record for event "i"
    pa  = eve.getParticles()        # particle information
    par = PromcUtil.getParticleDList(f.getHeader(),pa,1,10,3)
    clust.buildJets(par)            # build anti-kT jets
    jets = clust.getJetsSorted()    # get a list with sorted jets
    if (len(jets)>0):
        print "pT of a leading jet =",jets[0].perp()," GeV"
\end{lstlisting}

The method {\tt getParticleDList} extracts the list of particles that pass the specified
transverse momentum and pseudo-rapidity cuts.
Each particle in the list is represented by an instance of {\tt ParticleD}, which is
a simple class with cached kinematic variables for fast processing.
The third argument of {\tt getParticleDList} is the status code for particles to be accepted (for final state particles, use {\tt status=1}), the fourth argument is the
minimum $p_\mathrm{T}$, and the fifth argument is the maximum pseudo-rapidity
of the returned particles.

In addition, one can construct jets by accumulating {\tt ParticleD}s in a standard Java {\tt ArrayList} according to more refined selection criteria, and pass this list to the {\tt buildJets} method.

\subsection{Histogramming and canvases}

The validation framework also includes libraries necessary to fill 1D, 2D, and profile histograms and show them on 2D and 3D graphical canvases.
The histogramming package is based on the \texttt{FreeHep} library~\cite{freehep}, and the graphical canvases  adopted
from the community edition of \texttt{DMelt}~\cite{dmelt} (formerly, jHepWork \cite{Chekanov:1261772}).
The main histogram classes for 1D and 2D histograms are {\tt jhplot.H1D} and {\tt jhplot.H2D}, while
the most commonly used graphical canvases are {\tt jhplot.HPlot} from \texttt{DMelt}.
A detailed description of the graphical canvases and histogramming packages is beyond the scope of this paper.

\section{Working with reconstructed events}

The validation of the reconstructed events stored in the form of LCIO \cite{lcio} files can be done in a similar manner.
In the case of the LCIO files, one can check distributions of calorimeters hits, clusters, tracks after full detector simulation and reconstruction, 
as well as to verify truth-level Monte Carlo information which is included from the original ProMC files.
The LCIO files are readable and writable in Java, therefore, Jython scripts can be constructed
using the same histogramming packages and plotting canvases as in the case of the ProMC files.
The main difference compared to the truth-level validation is that the LCIO files should be processed using the Jas4pp Java package 
\cite{jas4pp}. It
includes not only the plotting libraries,  but also the LCSIM \cite{lcsim} and LCIO Java packages 
which are required to access
the information on reconstructed objects after full detector simulation. 

\section{Conclusion}

This paper describes a validation framework used to analyze MC event files stored in the HepSim repository.
It includes all basic software components to reconstruct cross sections, kinematic distributions and reconstructed objects 
(tracks, calorimeter clusters etc.) needed to
check the validity of Monte Carlo simulation results.
The framework uses Python (Jython) as the primary language and provides a cross-platform HEP analysis environment for the
end-users while reducing maintenance. It is fully self-contained and it does not depend on any OS-specific libraries besides Java.
It has versatility to work on all major platforms and heterogeneous computer clusters, can be used 
together with Java Web Start technology, and allows the usage of the Java programming language 
and several scripting languages implemented on
the Java platform,  such as Python, Ruby, Groovy, and BeanShell.
Some script examples for validation of generated and reconstructed MC events  
in the ProMC and LCIO file formats are available from the HepSim web page.

\section*{Acknowledgments}
This research was performed using resources provided by the Open Science Grid,
which is supported by the National Science Foundation and the U.S. Department of Energy's Office of Science.
We gratefully acknowledge the computing resources provided on Blues,
a high-performance computing cluster operated by the Laboratory Computing Resource Center at Argonne National Laboratory.
Argonne National Laboratory's work was supported by the U.S. Department of Energy, Office of Science 
under contract DE-AC02-06CH11357.

\section*{References}

\bibliographystyle{elsarticle-num}
\def\bibname{\Large\bf References}
\def\refname{\Large\bf References}
\pagestyle{plain}
\bibliography{biblio}

\end{document}